\documentclass[aps, prb,amsmath, amssymb, reprint]{revtex4-1}

\usepackage{amsmath} 
\usepackage{amssymb}
\usepackage{graphicx}
\usepackage{amsfonts}
\usepackage{dcolumn}% Align table columns on decimal point
\usepackage{bm}% bold math
\usepackage{color}
\usepackage{natbib} %sty file for Bibtex

\begin{document}

\title{Influence of dynamical decoupling sequences with finite-width pulses
on quantum sensing for AC magnetometry}

\author{Toyofumi Ishikawa}
 \email{toyo-ishikawa@aist.go.jp}
 \affiliation{National Institute of Advanced Industrial Science
and Technology, Tsukuba,  305-8565, Japan}%Lines break automatically or can be forced with \\

\author{Akio Yoshizawa}
 \affiliation{National Institute of Advanced Industrial Science
and Technology, Tsukuba,  305-8565, Japan}

\author{Yasunori Mawatari}
 \affiliation{National Institute of Advanced Industrial Science
and Technology, Tsukuba,  305-8565, Japan}

\author{Satoshi Kashiwaya}
 \affiliation{Department of Applied Physics, Nagoya University, Nagoya 464-8603, Japan}

\author{Hideyuki Watanabe}
 \affiliation{National Institute of Advanced Industrial Science
and Technology, Tsukuba,  305-8565, Japan}

\date{\today}% It is always \today, today,
             %  but any date may be explicitly specified

\begin{abstract}
Dynamical decoupling sequences with multiple pulses 
can be considered to exhibit filter functions for the time evolution of a qubit superposition state. 
They contribute to the improvement of coherence time
and qubit-phase accumulation due to a time-varying field and can thus achieve high-frequency-resolution spectroscopy.
Such behaviors find useful application in highly sensitive detection based on qubits
for various external fields such as a magnetic field.
Hence, decoupling sequences are indispensable tools for quantum sensing.
In this study, we experimentally and theoretically investigated 
the effects of finite-width pulses in the sequences 
on AC magnetometry utilizing nitrogen-vacancy centers in an isotopically-controlled diamond. 
We revealed that the finite pulse widths cause a deviation of the optimum time 
to acquire the largest phase accumulation due to the sensing field from 
that expected by filter functions neglecting the pulse widths, 
even if the widths are considerably shorter than the time period of the sensing field.
Moreover, we experimentally demonstrated that 
the deviation can be corrected by an appropriate time--frequency conversion.
Our results provide a guideline for the detection of an AC field with an accurate 
frequency and linewidth in quantum sensing with multiple-pulse sequences.
\end{abstract}

\maketitle

\section{Introduction}
The phase of a superposition state of a qubit is sensitive to
various fields such as a magnetic field. 
Recently, qubits have attracted considerable attention as highly sensitive sensors 
called quantum sensors,\cite{RevModPhys.89.035002, 
doi:10.1063/1.5011231}
which can be utilized for
a wide range of applications such as a magnetometry,\cite{Taylor2008, Balasubramanian2008, 
doi:10.1146/annurev-physchem-040513-103659,0034-4885-77-5-056503}
thermometry,\cite{Kucsko2013, Toyli8417, PhysRevB.91.155404}
and electrometry.\cite{Dolde2011, PhysRevA.95.053417}
Dynamical decoupling sequences with multiple pulses are broadly used 
as fundamental tools of not only quantum information processing 
but also quantum sensing.
They can improve the coherence time of a quantum sensor,\cite{Bar-Gill2013}
and thus its frequency resolution and sensitivity.\cite{PhysRevB.86.045214}
Moreover, the decoupling sequences possess filter functions for qubit-superposition phases,
rendering the qubits sensitive to the fields with their time periods matched to that of the filters. 
$\pi$ pulses in the sequences reverse time evolutions of qubit states and
modulate the filters, which enable arrangement of their properties. 
Such techniques have greatly advanced quantum sensing
and have led to the realization of time-correlation spectroscopy\cite{PhysRevB.84.104301, doi:10.1063/1.3497004, Laraoui2013} 
and measurements with ultra-high frequency resolution 
that is sufficient to detect chemical shifts via nuclear magnetic resonance signals.\cite{Schmitt832, Boss837, Glenn2018}

Characterizations of the dynamical decoupling sequences are generally modeled and discussed 
using filter functions with infinitely narrow pulses.\cite{PhysRevB.77.174509,Taylor2008, PhysRevLett.106.080802} 
In practice, however, the pulses have finite widths, raising the need to modify the filter functions.
In this study, we propose a model of finite-pulse-width effects on 
AC magnetometry by utilizing multiple-pulse sequences;
in addition, we experimentally verify the model using nitrogen-vacancy (NV) centers in a diamond.

\section{Model of Finite-pulse-width effects on AC magnetometry}
\subsection{Filter function with finite-width pulses}
\begin{figure}[bhtp]
	\begin{center}
		\includegraphics[keepaspectratio, width = 8.5cm]{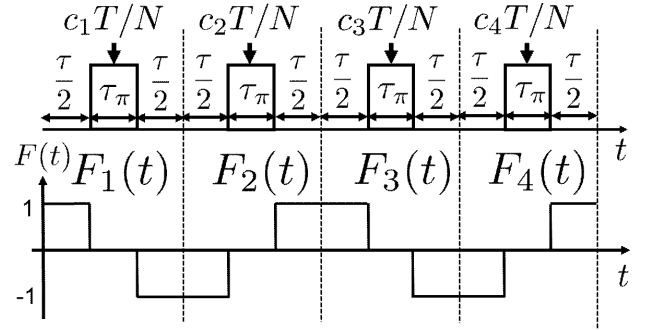}
	\end{center}
	\caption{A decoupling pulse sequence (upper figure)  and 
				a filter function  (lower figure) with $N$ finite-width pulses.
				$\tau_{\pi}$ is the pulse width, and $\tau$ is the free precession time between two pulses.
				Total time comprising the precession time and the width is $T=N ( \tau + \tau_{\pi} )$. 
				The center time of the j-th $\pi$ pulse is given by $c_{j} T/N,~c_j = (2j-1)/2$.
				The sequence with $N = 4$ is shown as an example.}
	\label{fig:FilterFunc}
\end{figure}

Dynamical decoupling sequences with multiple pulses 
are studied well previously.\cite{PhysRevB.77.174509, PhysRevLett.106.080802}  
Their behavior is characterized by filter functions $F(t) = \pm 1$, 
where the sign changes whenever a $\pi$ pulse is applied.
If the pulse width is neglected, 
the filter function shows a square wave form, 
where the modulation timing can be controlled by $\pi$ pulses.
However, when a finite pulse width is considered, the filter function must be modified.
Assuming that a qubit-control field generating $\pi$ pulses is significantly stronger than the sensing field,
the rotation axis of the qubit precession is parallel to the vector superposition of both fields,
which is approximately in the control-field direction. 
Therefore, phase accumulation due to the sensing field cannot occur during the generation of $\pi$ pulses,
and the filter function can be represented as shown in Fig. \ref{fig:FilterFunc}.
Here, $N$ is the number of $\pi$ pulses,
$\tau_{\pi}$ is the pulse width,
$\tau$ is the free precession time between two pulses, 
and total time $T=N ( \tau + \tau_{\pi} )$.
The center time of the j-th $\pi$ pulse is obtained by $c_{j} T/N,~c_j = (2j-1)/2$.

In our theoretical model, the filter function with finite-width $\pi$ pulses is given by
\begin{eqnarray}
F(t) &=& \sum_{j=1}^{N} F_j(t), 
\label{eq:Filter_time} \\
F_{j}(t) &=&
\left \{
\begin{array}{ll}
(-1)^{j-1} & ( (j-1)\frac{T}{N} \le t < c_j\frac{T}{N} - \frac{\tau_{\pi}}{2})\\
(-1)^j & ( c_j\frac{T}{N} + \frac{\tau_{\pi}}{2} \le t < j \frac{T}{N} )\\
0 & (\mathrm{otherwise})
\end{array} \right.
.
\end{eqnarray}
Through the Fourier transformation of the filter function, we have
\begin{eqnarray}
\label{eq:Filter_omega}
F(\omega) &=& \int_{-\infty}^{\infty} dt~F(t) e^{-i\omega t} \nonumber \\ 
&=&
\frac{1}{i\omega}
\left [
1 + (-1)^{N+1} e^{-i \omega T}   
\right .
\nonumber \\
&~& 
+
\left .
2 \sum_{j=1}^{N} (-1)^j e^{-i \omega c_j T/N }
\cos{ \frac{\omega \tau_{\pi}}{2} }
\right ].
\end{eqnarray}
Michael {\it et al.}\cite{PhysRevA.79.062324}
reported a similar derivation for the field of quantum information by using an ion trap.
To examine the finite-pulse-width effects on AC magnetometry, 
we calculate the Fourier component in the sequence with $N = 1$, as follows: 
\begin{eqnarray}
\label{eq:FilterEcho}
F(\omega) &=&
\frac{i 4 e^{- i \omega [ \tau (1 + \alpha) ]/2 } }{\omega} \nonumber \\
&~& \times
\sin  \left( \frac{\omega \tau}{4}  \right )
\sin \left [ \frac{\omega \tau (1 + 2 \alpha )}{4} \right ] ,
\end{eqnarray}
where $\alpha = \tau_{\pi}/\tau$.
The above equation corresponds to a filter function for a Hahn-echo sequence.
For the sequence with even $N$ pulses,
the Fourier component is given by
\begin{eqnarray}
\label{eq:FilterCP}
F(\omega) &=&
N \tau (1 + \alpha) 
e^{- i \omega N \tau (1+\alpha)/2}
\nonumber \\
&~&\times
\left \{
1 - \frac { \cos (\alpha \omega \tau / 2) }
{\cos [ \omega \tau (1 + \alpha) /2 ] } 
\right \}  
\nonumber \\
&~&\times
\frac{
\sin  \left [ \omega N \tau (1 + \alpha) /2  \right ]
}
{\omega N \tau (1 + \alpha)/2 },
\end{eqnarray}
which is a filter function for Carr-Purcell (CP) type sequences.\cite{PhysRev.94.630}

\subsection{Phase accumulation induced by an AC magnetic field}
We assume a single-tone AC magnetic field given by
\begin{equation}
\label{eq:ACfield}
B(t) = B_{ac} \cos{ 
\left (
\omega_{ac}t + \phi_{ac}
\right ).
}
\end{equation}
$\omega_{ac} = 2\pi f_{ac}$, where $f_{ac}$ is the field frequency.
$B_{ac}$ is the field strength
and $\phi_{ac}$ is the initial phase at the start of the decoupling sequences.
A quantum sensor acquires phase accumulation given by
\begin{eqnarray}
\label{eq:PhaseAccumulation}
\Phi &=& \int_{-\infty}^{\infty} dt F(t) \gamma_{NV} B(t) \nonumber \\
&=&
\gamma_{NV} B_{ac} \left .
\Re \{ e^{-i\phi_{ac}} F(\omega_{ac}) \}. \right.
\end{eqnarray}
$\gamma_{NV} $ is the gyromagnetic ratio of the quantum sensor.
In this study, we use NV centers in a diamond as quantum sensors.
Therefore, $\gamma_{NV}/{2\pi} = 28.03~\mathrm{GHz/T}.$
According to Eqs. (\ref{eq:FilterEcho}) and (\ref{eq:FilterCP}),
phase accumulation can be given as follows, 
on the basis of the Hahn-echo and the CP-type sequences:
\begin{eqnarray}
\Phi_{\mathrm{Echo}} &=&
\frac{ 4 \gamma_{NV} B_{ac} }{\omega_{ac} }
\sin \left [ \frac{\omega_{ac} \tau (1 + \alpha )}{2} + \phi_{ac} \right ]  \nonumber \\
&~& \times
\sin  \left( \frac{\omega_{ac} \tau}{4}  \right ) 
\sin \left [ \frac{\omega_{ac} \tau (1 + 2 \alpha )}{4} \right ] ,  
\label{eq:PhaseAccEcho}
\\
\Phi_{\mathrm{CP}} &=&
\gamma_{NV}B_{ac} 
\cos \left [
\frac{\omega_{ac} N \tau (1 + \alpha)}{2} + \phi_{ac}  
\right ] \nonumber \\
&~& 
\times
N \tau (1 + \alpha)
\left \{
1 - \frac { \cos (\alpha \omega_{ac} \tau / 2) }
{\cos [ \omega_{ac} \tau (1 + \alpha) /2 ] } 
\right \}
\nonumber \\
&~& \times
\frac{
\sin  \left [ \omega_{ac} N \tau (1 + \alpha) / 2  \right ]
}
{\omega_{ac} N \tau (1 + \alpha)/2 }. 
\label{eq:PhaseAccCP}
\end{eqnarray}

Commonly, a $\pi/2$ pulse is applied to an NV quantum sensor 
before the dynamical decoupling sequence, 
which creates the superposition state.
Furthermore, additional $\pi/2$ pulse is applied after the sequence, 
which enables optical detection of phase accumulation from NV centers.
When both $\pi/2$ pulses are in-phase, 
the signal of the AC magnetometry is proportional to $\cos{\Phi}$.
When the final $\pi/2$ pulse that is in quadrature with the other $\pi/2$ pulse is applied,
the magnetometry signals become proportional to $\sin{\Phi}$. 
  
\section{Experiments}
\subsection{Methods}
In this study, we used 
an ensemble of  NV centers ($\mathrm{[NV]}\approx 3 \times 10^{14}~\mathrm{cm}^{-3} $) as a quantum sensor. 
The NV centers were produced by a procedure reported in Ref. \onlinecite{7466817}, with some modification.
Briefly, a $^{12}$C diamond film was prepared 
by microwave-plasma-assisted chemical vapor deposition 
from isotopically enriched $\mathrm{ ^{12}C H_{4}}$  ($> 99.999~\%$ for $\mathrm{ ^{12}C} $) 
and $\mathrm{H_2}$ mixed gas.
The thickness of the diamond film is $50~\mathrm{nm}$.
We used a reactant gas with nitrogen to carbon ratio $\mathrm{N/C}=1.75~\%$ during the growth, 
and photolithography was used instead of electron beam lithography. 
We evaluated coherence and relaxation time of the ensemble:
$T_2 = 74.0 \pm 2.7 ~\mathrm{\mu s}$ and $T_1 = 7.04 \pm 0.27~\mathrm{ms}$.

The AC magnetometry measurements were performed in this study 
using a home-built laser scanning microscope with a 0.85 numerical aperture objective (Olympus LCPLFLN100x LCD).
Our microwave setup consists of a microwave source (QuickSyn Synthesizers FSW-0020)
and a quadrature hybrid coupler (Marki QH-0R714 Quadrature Hybrid Coupler)
with each output connected to a microwave switch (Mini-Circuits ZYSWA-2-500DR) 
for generating either in-phase or quadrature-phase microwave pulses.
After each switch, both output paths were combined, 
and then amplified by Mini-Circuits ZHL-16W-43+, followed by a microwave antenna.\cite{doi:10.1063/1.4952418} 
A commercial neodymium permanent magnet generated a static magnetic field of $\sim 4~\mathrm{mT}$
with its direction parallel to a $\left \langle 111 \right \rangle$ axis of NV centers.
An AC magnetic field was applied to the NV ensemble by a home-made coil 
connected to an arbitrary function generator (Tektronix AFG3252) 
and synchronized with dynamical-decoupling sequences. 
We canceled common-mode noises of our measurements based on dynamical decoupling sequences
by using a $3\pi/2$ pulse instead of a $-\pi/2$ pulse.\cite{Bar-Gill2013}

\subsection{Results and Discussions}

\begin{figure}[tbhp]
	\begin{center}
		\includegraphics[keepaspectratio,, width = 8.4cm]{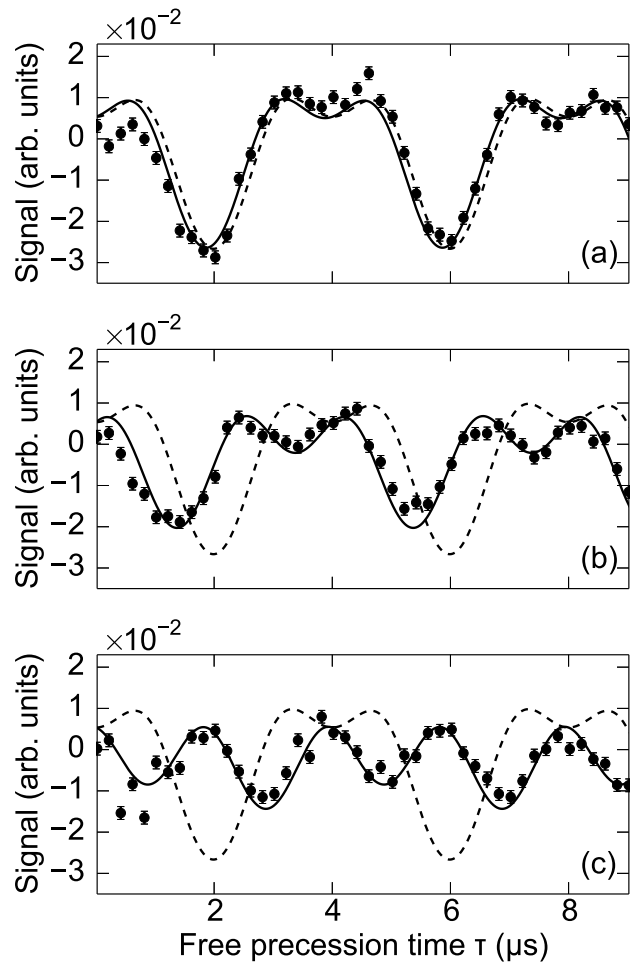}
	\end{center}
	\caption{AC magnetometry using a Hahn-echo sequence.
				Error bars in the figure indicate the standard deviation of photon shot noises.
				The frequency of the sensing field was set at $500~\mathrm{kHz}$.
				Its strength and phase were evaluated 
				by $\sin \Phi_{\mathrm{Echo}}$
				fitted to the result in the case of $\tau_{\pi} = 124~\mathrm{ns}$,
				indicated by the solid line in (a). 
				Detail procedure of the fitting is explained in Appendix. 
				(b) and (c) show the AC magnetometry
				with $\tau_{\pi} = 622~\mathrm{ns}$ and $1120~\mathrm{ns}$, respectively.
				The solid lines in (b) and (c) were obtained by $\sin \Phi_{\mathrm{Echo}}$, 
				where AC-field parameters were estimated from the result shown in (a) 
				and each $\tau_{\pi}$ were assigned. 
				The dashed line in each figure shows the theoretical plot without considering the pulse width,
				which is plotted by $\sin \Phi_{\mathrm{Echo}}$ where $\tau_{\pi} = 0$, 
				i.e., $\alpha = 0$.}
	\label{fig:HahnEcho}
\end{figure}

To examine the effects of the finite pulse width, 
we performed AC magnetometry using a Hahn-echo sequence ($N = 1$), 
i.e., $(\pi/2)_x$ -- $\tau/2$ -- $(\pi)_y$ -- $\tau/2$ -- $(\pi/2)_y$. 
A sensing field with 
$f_{ac} = 500 ~\mathrm{kHz}$
started subsequent to the first $(\pi/2)_x$ pulse, 
and the field was synchronized to the Hahn-echo sequence.
Figure \ref{fig:HahnEcho}(a) shows 
the AC magnetometry using the Hahn-echo sequence with $\tau_{\pi} = 124~\mathrm{ns}$.
The solid line in Fig. \ref{fig:HahnEcho}(a) indicates
the result fitted according to $\sin \Phi_{\mathrm{Echo}}$:
$B_{ac} = 2.75\pm0.13~\mathrm{\mu T}$ and
$\phi_{ac} =91.4 \pm 2.3~\mathrm{deg.}$
Details of our fitting are explained in Appendix.  
The largest phase accumulation was observed 
when the free precession time in the echo sequence was matched 
with the time period of the sensing field, i.e., 
$\tau \approx (2k + 1)T_{ac},~ \left ( k  = 0, 1, 2, \cdots, ~T_{ac}=1/f_{ac} = 2~\mathrm{\mu s} \right ) $.
This feature is consistent with the expectations 
from the filter functions without considering finite-width pulses,
as indicated by the dashed line in Fig. \ref{fig:HahnEcho}(a).
By contrast, 
as the pulse width increases,  
the AC magnetometry signals become increasingly different from that shown in Fig. \ref{fig:HahnEcho}(a).
Figures \ref{fig:HahnEcho}(b) and \ref{fig:HahnEcho}(c) show the results 
of AC magnetometry using the echo sequence 
with $\tau_{\pi} = 622~\mathrm{ns}$ 
and  $1120 ~\mathrm{ns}$, 
which are equivalent to $5\pi$ and $9\pi$.
We observed finite-width-pulse-induced deviations 
of the phase-accumulated times from Fig. \ref{fig:HahnEcho}(a) 
and degradation of the AC magnetometry signals.
The solid lines in Figs. \ref{fig:HahnEcho}(b) and \ref{fig:HahnEcho}(c) were
plotted using AC magnetometry signals obtained with 
$\sin{\Phi}$ with $\Phi = \Phi_{\mathrm{Echo}}$ [Eq. (\ref{eq:PhaseAccEcho})].
The solid lines in Figs. \ref{fig:HahnEcho}(b) and \ref{fig:HahnEcho}(c) 
obtained with the AC-field parameters evaluated from the result 
shown in Fig. \ref{fig:HahnEcho}(a)  and each $\tau_{\pi}$ assigned to Eq. (\ref{eq:PhaseAccEcho})
show consistency with the experimental results.

\begin{figure}[tbhp]
	\begin{center}
		\includegraphics[keepaspectratio,, width = 8.4cm]{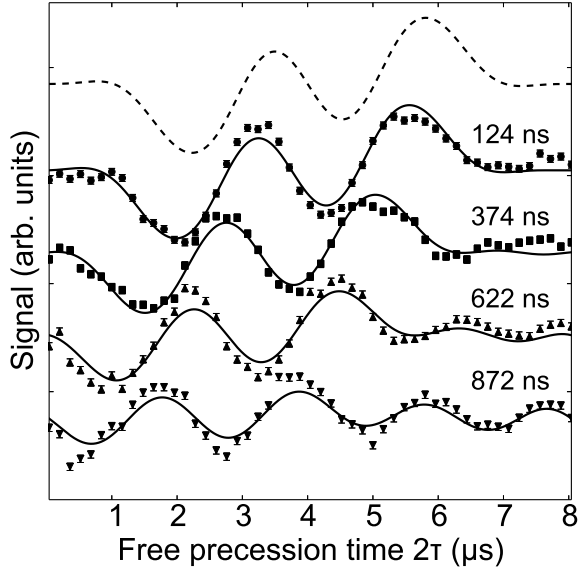}
	\end{center}
	\caption{AC magnetometry using a CPMG-2 sequence.
				The frequency of the sensing field was set at $500~\mathrm{kHz}$.
				Its strength and phase were evaluated by $\sin \Phi_{\mathrm{CP}}$ 
				fitted to the result with $\tau_{\pi} = 124~\mathrm{ns}$.
				The fitting result is indicated by the solid line 
				in the case of $\tau_{\pi} = 124~\mathrm{ns}$.
				These AC-field parameters were used to obtain the solid lines plotted by 
				$\sin \Phi_{\mathrm{CP}}$  for the results with
  				$\tau_{\pi} = 374~\mathrm{ns}$, 
				$622~\mathrm{ns}$, and $872~\mathrm{ns}$, 
				which were equivalent to $3\pi$, $5\pi$, and $7\pi$ pulses.
				The solid lines reproduce the results well.
				The dashed line shows the theoretical plot without considering the pulse width,
				which is plotted by $\sin \Phi_{\mathrm{CP}}$ where $\tau_{\pi} = 0$,
				i.e., $\alpha = 0$.
				}
	\label{fig:CPMG}
\end{figure}

Figure \ref{fig:CPMG} shows the AC magnetometry results based 
on a Carr-Purcell-Meiboom-Gill sequence\cite{doi:10.1063/1.1716296} with $N = 2$ (CPMG-2), 
i.e., $(\pi/2)_x$ -- $\tau/2$ -- $(\pi)_y$ -- $\tau$ -- $(\pi)_y$ -- $\tau/2$ -- $(\pi/2)_y$,
which is classified as the CP-type sequence.
As described in Appendix, $T_2(N=2)=94.7 \pm 3.5~\mathrm{\mu s}$.
The sensing field strength and frequency were the same as those for the echo-based magnetometry,
and we subtracted 90 degree from the AC phase of our Hahn-echo measurements.  
According to $\sin \Phi_{\mathrm{CP}}$ given by Eq. (\ref{eq:PhaseAccCP}), 
fitting the experimental result with $\tau_{\pi} = 124~\mathrm{ns}$ 
yielded $B_{ac} = 2.42\pm0.09~\mathrm{\mu T}$ and 
$\phi_{ac} =-2.53 \pm 2.05~\mathrm{deg.}$
We assigned the evaluated AC field parameters, $\tau_{\pi} =374~\mathrm{ns}$,
 $622~\mathrm{ns}$, and  $872~\mathrm{ns}$ 
to Eq. (\ref{eq:PhaseAccCP}), 
which reproduced our experimental results well.

\begin{figure}[thbp]
	\begin{center}
		\includegraphics[keepaspectratio,, width = 8.4cm]{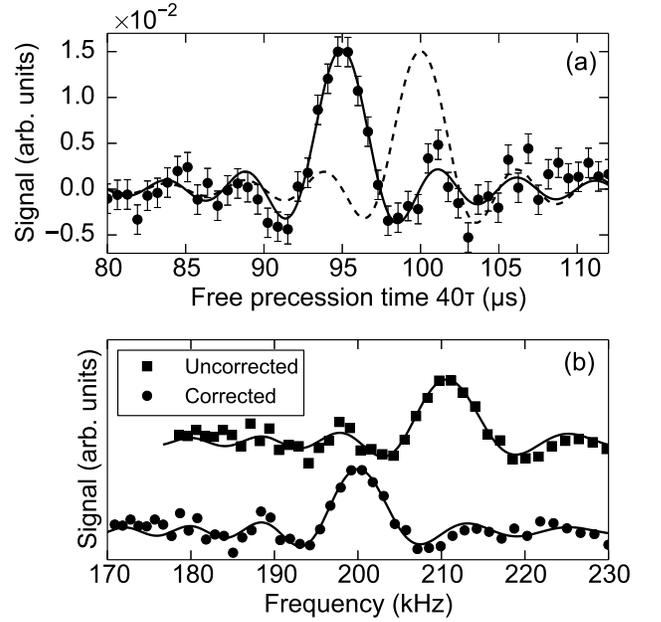}
	\end{center}
	\caption{(a) AC magnetometry using an XY8-5 sequence ($N = 40$).
				XY-series sequences have the same filter functions as that of the CP-type ones.
				Error bars are obtained from the standard deviation of photon shot noises.
				The sensing field frequency was set at $200~\mathrm{kHz}$.
				The solid line shows the fitting result according to $\sin \Phi_{\mathrm{CP}}$, 
				whereas the dashed line in (a) indicates $\sin \Phi_{\mathrm{CP}}$ with $\alpha = 0$.
				(b) XY8-5 results plotted in a frequency range. 
				Frequency conversion with $f =(2\tau)^{-1}$ causes 
				a $10~\mathrm{kHz}$ shift from the set frequency.
				By contrast,  frequency conversion with 
				$f = \left [ 2(\tau + \tau_{\pi}) \right ]^{-1}$
				corrects the deviation.
				}
	\label{fig:XY8}
\end{figure}
　
Finally, AC magnetometry results 
based on XY-series sequences with pulse errors suppressed\cite{GULLION1990479}
were obtained.
Those sequences are classified as the CP-type decoupling sequences 
and are widely used for NV-based quantum sensing.
\cite{PhysRevLett.106.080802, Staudacher561, 
doi:10.1063/1.4862749, Muller2014,  Rugar2014,DeVience2015, Haberle2015, Glenn2018, 
Staudacher2015, Wood2017}
Figure \ref{fig:XY8}(a) shows the AC magnetometry using the XY8-5 sequence.
The sensing field frequency was set at $200~\mathrm{kHz}$, 
its estimated strength was $45.6 \pm 2.9 ~\mathrm{nT}$, 
and $\phi_{ac} = 181.2 \pm 3.3~\mathrm{deg.}$ 
$\tau_{\pi}$ was $126~\mathrm{ns}$.
The dashed line in Fig. \ref{fig:XY8}(a) 
indicates the theoretical plot with infinitely narrow pulses ($\alpha = 0$).
Although the optimum precession time for the phase accumulation was 
expected to be much longer than the widths, 
implying that $2\tau \sim 1/f_{ac} = 5~\mathrm{\mu s} \gg \tau_{\pi}$ and $\alpha \ll 1$,
we observed the deviation of the peak from
that indicated by $\sin \Phi_{\mathrm{CP}}$ with $\alpha = 0$, 
as seen in Fig. \ref{fig:XY8}(a).
If $\cos{\left ( \alpha \omega_{ac}\tau / 2  \right ) }\sim 1$, 
Eq. (\ref{eq:FilterCP}) at the neighborhood of $\omega_{ac}$ is approximately equal to 
\begin{eqnarray}
\left | F(\omega) \right | 
&\approx&
N \tau (1 + \alpha)
\left \{
1 -
\sec [ \omega \tau (1 + \alpha) /2 ] 
\right \}
\nonumber \\
&~& \times
\frac{
\sin  \left [ \omega N \tau (1 + \alpha) / 2  \right ]
}
{\omega N \tau (1 + \alpha)/2 }. 
\label{eq:ApproxPhaseAccCP}
\end{eqnarray}
In our experimental conditions, 
this approximation is valid because $  \alpha \omega_{ac}\tau/2 \approx 0.025~\mathrm{rad}$.
Equation (\ref{eq:ApproxPhaseAccCP}) is similar 
to the filter functions with infinitely narrow $\pi$ pulses,\cite{RevModPhys.89.035002, Taylor2008} 
where $\tau$ is replaced with $\tau(1+\alpha) = \tau + \tau_{\pi}$.
It implies that the frequency conversion 
using  $f = \left [ 2(\tau + \tau_{\pi}) \right ]^{-1}$  is available 
to correct the finite-pulse-width-induced deviation. 
Figure \ref{fig:XY8}(b) shows the frequency plot of the XY8-based AC magnetometry result.
Uncorrected experimental result plotted as a function of $f =(2\tau)^{-1}$ 
causes a $10~\mathrm{kHz}$ shift of the main peak from the set frequency.
In contrast,
in the corrected conversion, i.e., the data plotted as a function of $f = \left [ 2(\tau + \tau_{\pi}) \right ]^{-1}$, 
the main peak appears at 200 kHz, which is consistent with the set frequency in this experiment.
As shown in Fig. \ref{fig:XY8}, the AC magnetometry measurements 
using CP-type sequences with even pulses 
result in signals possessing a sinc function given by 
$ \mathrm{sinc} { \left [ \omega N \tau (1 + \alpha ) /2\right ] }$
and their linewidths $\approx 1/T$, where $T = N\tau (1 + \alpha)$.
However, an uncorrected conversion 
plotted as a function of $f =(2\tau)^{-1}$ neglects $\alpha$, 
which results in their linewidths in a broader frequency range than that of the corrected-frequency plots.
Therefore, 
finite-width pulses should be taken into account 
in order to evaluate accurate frequency and linewidth through AC magnetometry 
using dynamical decoupling sequences with multiple $\pi$ pulses.

\section{Conclusion}
We studied the effects of finite $\pi$-pulse widths on AC magnetometry 
theoretically and experimentally 
by using NV quantum sensors in an isotopically purified diamond film.
Finite pulse widths comparable to the time period of the sensing field 
induce deviations of the optimum free precession times
from the phase-accumulated times expected 
from the filter functions when finite-width-pulses are not taken into account.
Furthermore, AC magnetometry signals are degraded.
Therefore, high-frequency field detection requires pulses as narrow as possible
to evaluate optimum time and improve the signal-to-noise ratio.
Furthermore, long-time measurements with many $\pi$ pulses 
may result in evident finite-width-pulse-induced deviations,
even though the widths are much shorter than the time period of the sensing field.
Such deviations cause failure in the estimation of the optimum free precession time for phase accumulation, 
resulting in the degradation of signals of AC magnetometry performed at fixed $\tau$.
Therefore, to estimate the optimum free precession time 
and detect AC signals with accurate frequency and linewidth, 
quantum sensing for AC magnetometry 
based on decoupling sequences with multiple pulses 
must take finite-width pulses into account.

\section*{Acknowledgements}
We thank Kento Sasaki, Eisuke Abe, Junko Ishi-Hayase, and Kohei M. Itoh in Keio University 
for supporting the start-up of our optical and microwave-control systems.
We also acknowledge Hitoshi Sumiya from Sumitomo Electric Industries for supplying the diamond substrate
and Yoshikiyo Toyosaki from Correlated Electronics Group in AIST for the technical support in performing photolithography.
This work was supported in part by SENTAN.JST, 
Grant-in-Aid for Young Scientists (B) Grant Number JP17K14079,
and JSPS Kakenhi no. JP15H05853.

\section*{Appendix}

In this study, a common-mode-rejection method was used to obtain the results.\cite{Bar-Gill2013}
Therefore, the signals of AC magnetometry using the pulse sequences described in the main text are given by
\begin{equation}
\label{eq:Signal_q}
S_{Q} = (-1)^{n_y +1}\frac{1-r}{1+r} 
\exp{
\left [
- \left ( \frac{N\tau}{T_2(N)} \right ) ^ p
\right ]
}
\sin{\Phi} ,
\end{equation} 
where $r$ is the ratio of photon counts from the dark state to the bright state of the NV center,
$n_y$ is the number of $(\pi)_y$ pulses, 
and $T_2(N)$ indicates the coherence time of $N$-pulse decoupling sequences.
$\Phi$ is the phase accumulation due to the sensing field,
which is $\Phi_{\mathrm{Echo}}$ [Eq. (\ref{eq:PhaseAccEcho})] 
in the measurements using the Hahn-echo sequences 
and $\Phi_{\mathrm{CP}}$ [Eq. (\ref{eq:PhaseAccCP})] 
in the measurements using the CP-type sequences.

\begin{figure}[thbp]
	\begin{center}
		\includegraphics[keepaspectratio,, width = 8.4cm]{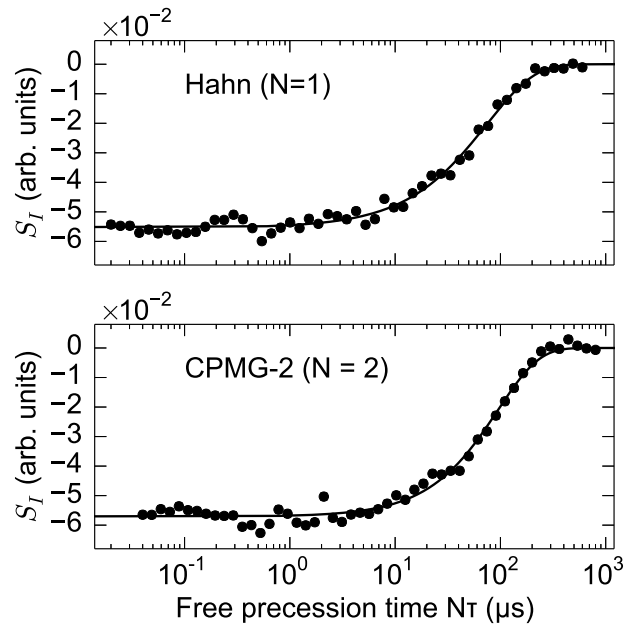}
	\end{center}
	\caption{Evaluations of coherence time of an ensemble of NV centers 
				using the Hahn-echo ($N = 1$) and CPMG-2 ($N=2$) sequences.
				We applied $(\pi)_y$ pulses for both sequences, i.e., $n_x = 0$ in Eq. (\ref{eq:Signal_I}).}
	\label{fig:Coherence}
\end{figure}

Using the Hahn-echo-sequence- and CPMG-sequence-based measurements without the sensing field, 
where the first and final $\pi/2$ pulses are in-phase, 
we evaluated the parameters $r$, $T_2(N)$, and $p$, because
the signals are given by
\begin{equation}
\label{eq:Signal_I}
S_{I} = (-1)^{n_x +1}\frac{1-r}{1+r} 
\exp{
\left [
- \left ( \frac{N\tau}{T_2(N)} \right ) ^ p
\right ]
},
\end{equation}
where $n_x$ means the number of $(\pi)_x$ pulses.
Figure \ref{fig:Coherence} shows the Hahn-echo ($N=1$) measurement result,
revealing $r = 0.895 \pm 0.001$, $T_2(N=1) = 74.0 \pm 2.7 ~\mathrm{\mu s}$,
and $p = 0.952 \pm 0.004$.
In the CPMG-2 ($N=2$) measurement result shown in Fig. \ref{fig:Coherence}, 
$r = 0.892 \pm 0.001$, $T_2(N=2) = 94.7 \pm 3.5 ~\mathrm{\mu s}$,
and $p = 1.11 \pm 0.06$.
These evaluated parameters and Eq. (\ref{eq:Signal_q}) were used 
for fitting the results described in the main text.

\newpage

%\bibliographystyle{apsrev4-1}	%.bst file is read. 
%\bibliography{bib}	%Bibliography database file. Here bib.bib file is read. 

%merlin.mbs apsrev4-1.bst 2010-07-25 4.21a (PWD, AO, DPC) hacked
%Control: key (0)
%Control: author (72) initials jnrlst
%Control: editor formatted (1) identically to author
%Control: production of article title (-1) disabled
%Control: page (0) single
%Control: year (1) truncated
%Control: production of eprint (0) enabled
%

\end{document}